\documentclass[aps,prl,reprint,longbibliography,superscriptaddress]{revtex4-2}
\usepackage{mathrsfs}
\usepackage{amsmath,gensymb}
\usepackage{amsfonts}
\usepackage{amssymb}
\usepackage{amsthm}
\usepackage{graphicx}
\usepackage{natbib}
\usepackage{color}
\usepackage{hyperref}
\usepackage{bm}
\usepackage[caption=false]{subfig}
\usepackage{verbatim}
\usepackage[normalem]{ulem}

\begin{document}
\title{Magnetic anisotropy of superconducting transition in S/AF heterostructures with spin-orbit coupling}

\author{G. A. Bobkov}
\affiliation{Moscow Institute of Physics and Technology, Dolgoprudny, 141700 Moscow region, Russia}

\author{I.V. Bobkova}
\affiliation{Moscow Institute of Physics and Technology, Dolgoprudny, 141700 Moscow region, Russia}
\affiliation{National Research University Higher School of Economics, 101000 Moscow, Russia}

\author{A.A. Golubov}
\affiliation{Faculty of Science and Technology and MESA$^+$ Institute for Nanotechnology, University of Twente, 7500 AE Enschede, The Netherlands}

\begin{abstract}
The influence of Rashba spin-orbit coupling (SOC) on superconducting correlations in thin-film superconductor/antiferromagnet (S/AF) structures with compensated interfaces is studied. A unique effect of anisotropic enhancement of proximity-induced triplet correlations by the SOC is predicted. It manifests itself in the anisotropy of the superconducting critical temperature $T_c$ with respect to orientation of the N\'eel vector relative to the S/AF interface, which is opposite to the behaviour of $T_c$ in superconductor/ferromagnet  structures. We show that the anisotropy is controlled by the chemical potential of the superconductor and, therefore, can be adjusted in (quasi)2D structures.  
\end{abstract}

\maketitle

{\it Introduction.}---
The interplay between superconductivity and ferromagnetism in thin film superconductor/ferromagnet (S/F) heterostructures  usually manifests itself as a change in superconductivity of the S layer due to the proximity to the F layer. The most well-known and studied effect is induced by the magnetic proximity triplet superconducting correlations\cite{Buzdin2005,Bergeret2005}.  Further studies \cite{Gorkov2001,Annunziata2012,Bergeret2013,Bergeret2014,Edelstein2003,Edelstein2003_JETPLett,Jacobsen2015,Mackenzie2003,Linder_review} have predicted and observed that spin-orbit coupling (SOC) in S/F bilayers can produce an anisotropic depairing effect on triplets. One of the manifestations of the anisotropic depairing is that the critical temperature $T_c$ of the bilayer depends on the orientation of the F layer magnetization with respect to the S/F interface\cite{Jacobsen2015,Ouassou2016,Simensen2018,Banerjee2018}. 

One of consequences of a SOC-driven modulation of superconductivity is the possibility for a reciprocal effect i.e., a reorientation of the F layer magnetization due to superconductivity\cite{Jonsen2019,Gonzalez-Ruano2020,Gonzalez-Ruano2021}. For sufficiently thin ferromagnetic layers, a
change from in-plane (IP) to out-of-plane (OOP) magnetization has been predicted \cite{Jonsen2019} and realised \cite{Gonzalez-Ruano2021} in
magnetic tunnel junctions. The possibility to control magnetic anisotropies using superconductivity - a key step
in designing future cryogenic magnetic memories and spintronics applications.

However, the finite net magnetization of ferromagnets presents a significant drawback
for applications in nanoscale devices. On the other hand, antiferromagnets (AFs) are magnetically ordered materials
with zero net magnetization and negligible stray fields,
as well as intrinsic high-frequency dynamics. Due to these advantages they are being actively studied as alternatives to ferromagnets for
spintronics applications\cite{Baltz2018,Jungwirth2016,Brataas2020}. For AF-based superconducting spintronics it is of crucial importance to study proximity effects in superconductor/antiferromagnet (S/AF) heterostructures \cite{Andersen2006,Enoksen2013,Bobkova2005,Andersen2005,Jonsen2021,Bell2003,Hubener2002,Wu2013,Seeger2021}. It has been reported that in S/AF structures superconducting triplet correlations also arise due to the proximity effect. Depending on the particular antiferromagnetic order, system geometry and the interface properties they can be of different types. In particular, if the S/AF interface possesses nonzero net magnetization (uncompensated interface), it induces Zeeman splitting and conventional triplet correlations in the adjacent superconductor\cite{Kamra2018}. The compensated S/AF interfaces also induce triplet correlations, but they are of the N\'eel type\cite{Bobkov2022,Bobkov2023}, that is their amplitude flips sign from one lattice site to the next, just
like the N\'eel spin order in antiferromagnets. For S/AF heterostructures with canted AFs the mixture of conventional and N\'eel triplet correlations has been predicted \cite{Chourasia2023}.

At the same time effects of SOC on the proximity effect in S/AF hybrids are much less explored. In particular, the anomalous phase shift in S/AF/S Josephson junctions with SOC\cite{Rabinovich2019}, anisotropy of the critical current \cite{Falch2022} and topological superconductivity in S/AF hybrids \cite{Lado2018} have been predicted, also the anisotropic magnetoresistance has been calculated\cite{Jakobsen2020}. Here we study anisotropic effect of Rashba SOC on triplets in S/AF thin film bilayers with fully compensated AFs. It is found that in addition to the anisotropic depairing of triplet correlations known in S/F hybrids, a unique effect of anisotropic enhancement of the triplets by the SOC occurs in the S/AF case. We unveil the physical mechanism of the effect and demonstrate that it can manifest itself in opposite trend in the anisotropy of the superconducting transition as compared to S/F heterostructures. Namely, in S/F thin film bilayers the critical temperature is higher for OOP magnetization orientation than for the IP magnetization\cite{Banerjee2018} due to the fact that 
SOC suppresses triplets oriented OOP more than
triplets oriented IP. Here we demonstrate the possibility of the opposite effect for S/AF thin-film bilayers with SOC, which occurs due to the anisotropic enhancement of triplets by SOC.

\begin{figure}[tb]
	\begin{center}
		\includegraphics[width=75mm]{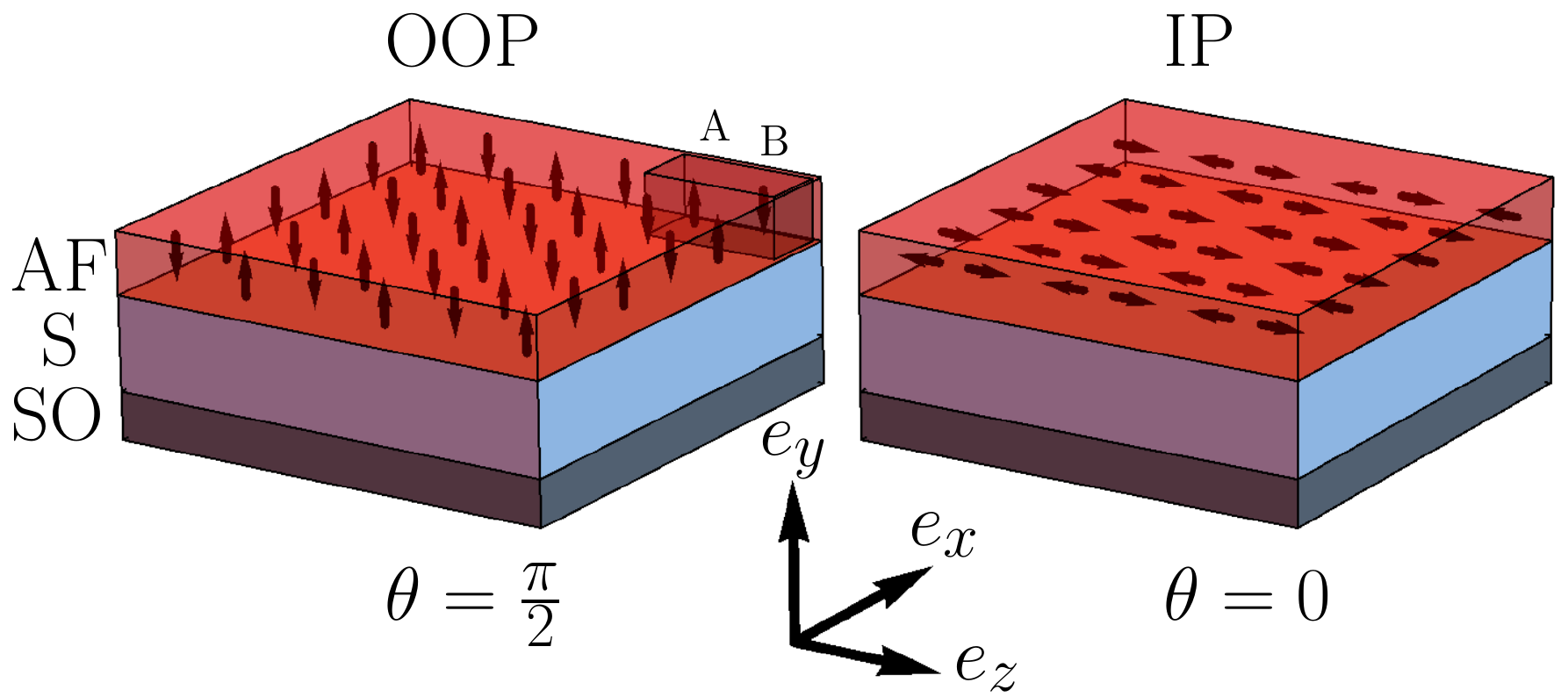}
		\caption{Sketch of the thin-film AF/S bilayer with SOC. The SOC is induced in the S layer by proximity to a heavy metal layer like Pt (shown as the SO layer). The SOC can also be due to inversion inversion symmetry breaking in the S film by itself. The N\'eel vector of the AF makes angle $\theta$ with the plane of the structure. $\theta=0$ corresponds to the IP and $\theta=\pi/2$ accounts for the  OOP orientations. Unit cell with two sites A and B is also shown.}
        \label{fig:setup}
	\end{center}
\end{figure}

{\it System and theoretical approach.}---We consider a thin-film S/AF bilayer, where the antiferromagnet is assumed to be an insulator, see Fig.~\ref{fig:setup}. The magnetism is staggered. We assume that the S/AF interface is fully compensated, that is the interface magnetization has zero average value. The sites in the superconductor are marked by the radius-vector $\bm i = (i_x,i_y, i_z)^T$, the interface is in the $(x,z)$-plane. The influence of the antiferromagnetic insulator on the superconductor is described by the  exchange field $\bm h_{\bm i} = \bm (-1)^{i_x+i_z} \bm h $ \cite{Kamra2018}.  The superconductor S is assumed to be homogeneous along the $y$-direction and  is described by the lattice Hamiltonian: 
\begin{align}
\hat H= - &t \sum \limits_{\langle \bm{i}\bm{j}\rangle,\sigma} \hat c_{\bm{i} \sigma}^{\dagger} \hat c_{\bm{j} \sigma} + \sum \limits_{\bm{i}} (\Delta_{\bm{i}} \hat c_{\bm{i}\uparrow}^{\dagger} \hat c_{\bm{i}\downarrow}^{\dagger} + H.c.) - \nonumber \\ 
&\mu \sum \limits_{\bm{i}, \sigma} \hat n_{\bm{i}\sigma}  +
 \sum \limits_{\bm{i},\alpha \beta} \hat c_{\bm{i}\alpha}^{\dagger} (\bm{h}_{\bm{i}} \bm{\sigma})_{\alpha \beta} \hat c_{\bm{i}\beta} + \nonumber \\ 
&i V_R \sum \limits_i (\hat c_{\bm i}^\dagger  \sigma_z \hat c_{\bm i+\bm e_x}-\hat c_{\bm i}^\dagger  \sigma_x  \hat c_{\bm i+\bm e_z}-H.c.),
\label{ham}
\end{align}
where $\langle \bm i \bm j \rangle $ means summation over the nearest neighbors, $\hat{c}_{\bm i \sigma}^{\dagger}(\hat{c}_{\bm i \sigma})$ is the creation (annihilation) operator for an electron with spin $\sigma$ at site $\bm i$. $t$ parameterizes the hopping between adjacent sites, $\Delta_{\bm{i}}$ accounts for on-site s-wave pairing, $\mu$ is the electron chemical potential, and the last term describes the Rashba SOC with Rashba constant $V_R$. $\hat n_{\bm i \sigma} = \hat c_{\bm i \sigma}^{\dagger} \hat c_{\bm i \sigma}$ is the particle number operator. $\bm e_k$ with $k=x,y,z$ are unit vectors along the corresponding axis. The lattice constant is denoted by $a$. Here we define Pauli matrices $\bm \sigma = (\sigma_x, \sigma_y, \sigma_z)^T$ in spin space.

\begin{figure}[tb]
	\begin{center}
		\includegraphics[width=65mm]{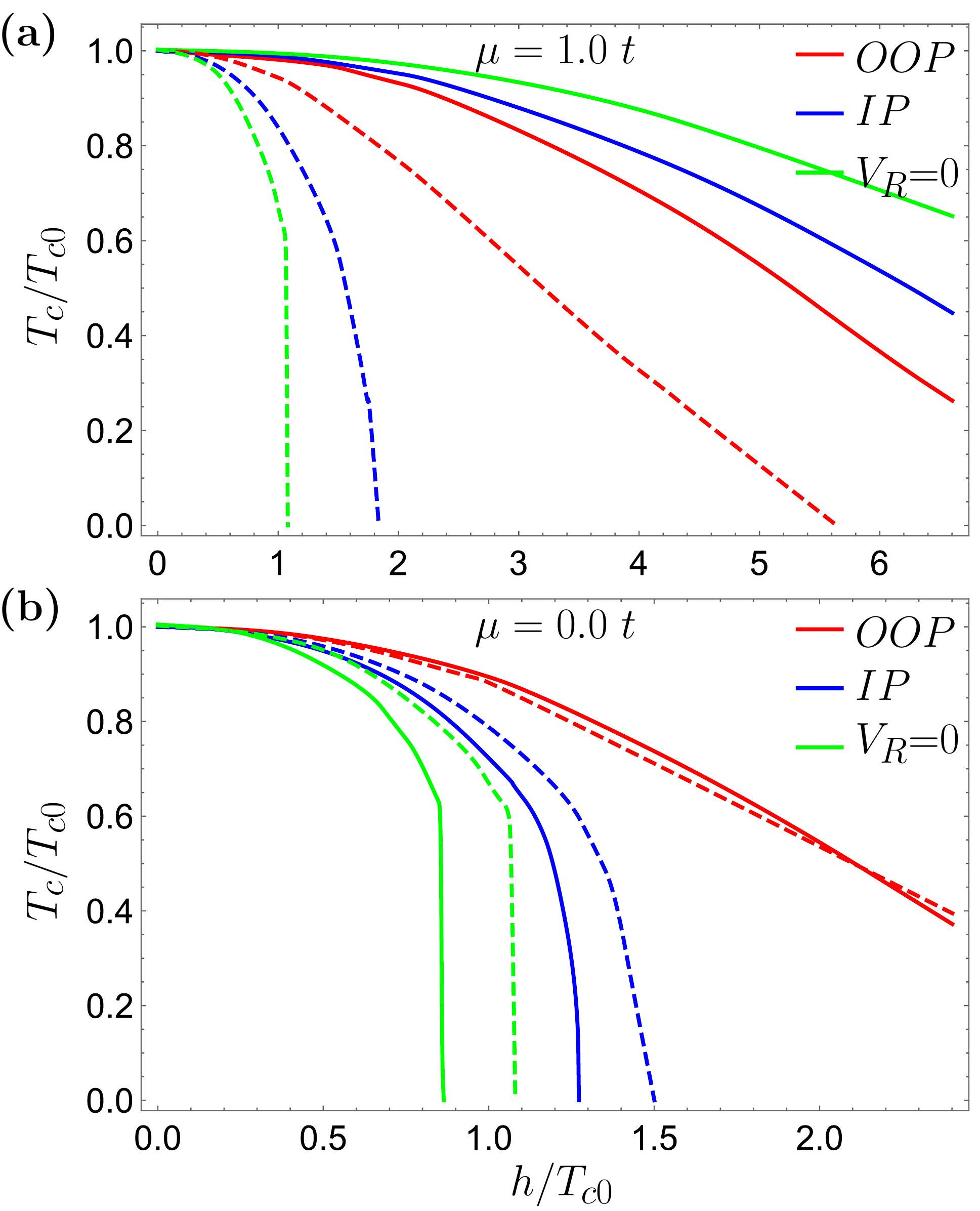}
		\caption{Critical temperature of S/AF (solid curves) and S/F bilayers (dashed) as a function of the induced exchange field $h$. Panels (a) and (b) correspond to $\mu=t$ and $\mu=0$, respectively. Green curves represent the results with no SOC, red and blue curves are for OOP and IP orientations, respectively and $V_R = 0.4t$. $T_{c0} = t/12$ for the both panels.}
        \label{fig:Tc}
	\end{center}
\end{figure}

{\it Anisotropy of triplets and $T_c$.}---The numerical calculations are performed in the formalism of the Gor'kov Green's functions in two-sublattice framework\cite{Bobkov2022,Bobkov2023}, generalized to take into account the SOC. Relegating technical details of the Green's functions formalism to the Supplemental material\cite{suppl}, here we present and discuss the dependencies $T_c(h)$ for S/AF structures with IP and OOP orientations of the N\'eel vector. They have been compared to $T_c(h)$ of the S/F system with the same absolute value of the induced exchange field $h$. The numerical results are presented in Fig.~\ref{fig:Tc}. First, it is seen that while for S/F heterostructures $T_c$ is always higher in the presence of SOC (dashed curves), for AF/S heterostructures the trends are opposite for large $\mu \gg T_{c0}$ and for small $\mu \lesssim T_{c0}$, where $T_{c0}=T_c(h=0)$. At small $\mu$ the behavior of $T_c$ is qualitatively similar to the case of S/F bilayers, and at large $\mu$ it is opposite - the presence of SOC {\it suppresses} $T_c$.

Furthermore, in the presence of SOC $T_c$ is anisotropic depending on the angle $\theta$ between the magnetization and the interface plane. For S/F heterostructures $T_c$ is always higher for OOP orientation (dashed curves) \cite{Jacobsen2015,Ouassou2016,Simensen2018,Banerjee2018}. At the same time for AF/S heterostructures the ratio between the values of $T_c$ for IP and OOP orientations is again opposite for large $\mu \gg T_{c0}$ and for small $\mu \lesssim T_{c0}$.
At $\mu \lesssim T_{c0}$ for S/AF heterostructures the ratio between $T_c$ of IP and OPP is the same as for the S/F case. It is explained by the fact that triplet superconducting correlations, induced by the proximity effect with the magnet are suppressed stronger for OOP configuration and, consequently, have a less damaging effect on the singlet superconductivity. At $\mu \gg T_{c0}$ the anisotropy of the critical temperature, that is the difference between the IP and OOP $T_c$ is opposite. This is due to the existence of a unique mechanism of {\it enhancement} of the N\'eel-type triplet correlations in S/AF structures, which is more effective for OOP orientation of the N\'eel vector.

The physical description of the both mechanisms is provided below. There is a crossover between the opposite anisotropy regimes at some intermediate value of $\mu \sim \pi T_{c0}$. For the chosen parameters of the the system this crossover value $\mu_c \approx 0.25 t \approx \pi T_{c0}$. It is interesting that the ratio between $\mu$ and superconduting energy scale $T_{c0}$ is crucial for very different aspects of the proximity physics of S/AF heterostructures. For example, this parameter controls the relative importance of different mechanisms of superconductivity suppression in S/AF hybrids \cite{Bobkov2023}. The superconductivity suppression is
dominated by the N\'eel triplets at $\mu \lesssim T_{c0}$. On the contrary, if $\mu \gg T_{c0}$ the superconductivity suppression is dominated by  nonmagnetic disorder. For S/AF heterostructures with canted AFs the opposite dependencies of $T_c$ on the canting angle were also predicted at small and large values of $\mu$ \cite{Chourasia2023}.

{\it Discussion of the mechanisms of $T_c$ anisotropy.}---Now we discuss the physical reasons of $T_c$ anisotropy in S/AF structures. In order to unveil them let us consider quasiclassical Eilenberger equations, developed recently for treating the proximity effect in S/AF heterostructures \cite{Bobkov2022} and their analytical solutions in the presence of Rashba SOC. The general form of the Eilenberger equation, generalized for treating the SOC, is provided in the Supplemental material\cite{suppl}.  In the vicinity of the critical temperature the Eilenberger equation can be linearized with respect to the anomalous Green's function, which for the problem under consideration can be written as $\check f = f_s \sigma_0 \rho_x + \bm f^i \bm \sigma \rho_i$. Here $f_s$ is its singlet component in spin space, $i=0,y,z$ and $\bm f^i$ is the vector triplet component in spin space, corresponding to the $0,y,z$ component in sublattice space. Recall that the $y$-component in sublattice space accounts for the on-site N\'eel-type  triplet correlations $\bm f^{AA} = -\bm f^{BB}$, while the $z(0)$-component describes nonlocal N\'eel (conventional) correlations $\bm f^{AB} = -(+) \bm f^{BA}$. The linearized equations take the form:
\begin{align}
&2i\omega f_s - 2 i\bm h \bm f^y = \Delta (g_N - \tilde g_N), \label{eq:eilenberger_singlet} \\
&2i\omega \bm f^0 + 2 i\bm h \times \bm f^z = \Delta (\bm g_N^0 - \tilde {\bm g}_N^0), \label{eq:eilenberger_0} \\
&\pm 2 i \mu \bm f^{y(z)} + 2 i \bm f^{z(y)} \times \bm h_R = \mp i \Delta (\bm g_N^{y(z)} + \tilde {\bm g}_N^{y(z)}), \label{eq:eilenberger_triplet} 
\end{align}
where $\bm h_R = (V_R/2ta) (\bm e_y \times \bm v_F)$ is the effective Rashba pseudomagnetic field seen by an electron moving along the trajectory determined by the Fermi velocity $\bm v_F$ and $\omega $ is the fermionic Matsubara frequency. The right-hand side contains the electron (hole) normal Green's function $\check g_N(\check {\tilde g}_N ) = g_N (\tilde { g}_N )\sigma_0 \rho_x + \bm g_N^i \bm \sigma (\tilde {\bm g}_N^i \bm \sigma)\rho_i$. The vector part of the on-site normal state quasiclassical Green's function accounting for the N\'eel order up to the leading order with respect to $(h, h_R)/|i\omega+\mu|$ takes the form:
\begin{eqnarray}
\bm g_N^y = -\frac{i \bm h {\rm sgn}\omega}{i\omega+\mu} - \frac{i [\bm h_R \times (\bm h \times \bm h_R)]}{(i\omega + \mu)^3}{\rm sgn}\omega \label{eq:normal_GFy} \\
\bm g_N^z = \frac{i \bm h_R \times \bm h}{(i\omega+\mu)^2}{\rm sgn}\omega .
\label{eq:normal_GFz}
\end{eqnarray}
The vector component $\bm g_N^y$ accounts for the N\'eel-type spin polarization of the on-site DOS $P_{\bm N}^A = -P_{\bm N}^B \equiv P_{\bm N}$ along the direction $\bm N$ ($|\bm N| = 1$) in spin space. 
\begin{eqnarray}
P_{\bm N}(\varepsilon) = 2N_F{\rm Re}[i \bm N \bm g_N^{R,y}(\varepsilon)],
\label{eq:dos}
\end{eqnarray} 
where $\bm g^{R,y}(\varepsilon) $ is obtained from $\bm g_N^y(\omega>0)$ by substitution $i\omega \to \varepsilon+i\delta$ and $N_F$ is the normal state DOS at the Fermi surface and at $h=0$.

It is important to compare Eqs.~(\ref{eq:eilenberger_singlet})-(\ref{eq:eilenberger_triplet}) to the analogous equations for S/F heterostructures. In this case the anomalous Green's functions has no N\'eel-type $\rho_{y(z)}$ components and the corresponding Eilenberger equations can be written without the sublattice structure, but we write them in terms of the same two-sublattices formalism in order to be directly compared with the AF case. In this case $\hat f_F = f_{F,s}\sigma_0 \rho_x + f_{F,0}\sigma_0 \rho_0 +\bm f_{F}^x \bm \sigma \rho_x + \bm f_{F}^0 \bm \sigma \rho_0$. It obeys the  following Eilenberger equations:
\begin{align}
&2i\omega f_{F,s} - 2 \bm h \bm f_F^x = 2 \Delta {\rm sgn}\omega,  \label{eq:eilenberger_singlet_F} \\
&2i\omega f_{F,0} - 2 \bm h \bm f_F^0 = 0,  \label{eq:eilenberger_singlet_0} \\
&i \omega \bm f_F^{x(0)} - \bm h f_{F,s(0)} + i \bm f_F^{0(x)} \times \bm h_R = 0 .  \label{eq:eilenberger_triplet_F} 
\end{align}
The most important difference between equations for triplet correlations for AF/S and F/S structures, Eqs.~(\ref{eq:eilenberger_triplet}) and (\ref{eq:eilenberger_triplet_F}) is that they contain different mechanisms of triplet generation. While in S/F hybrids the triplets are generated by the term $\bm h f_{F,s(0)}$ via the direct singlet-triplet conversion \cite{Buzdin2005}, such type of triplet generator is absent for AF/S structures. Instead, the triplets are generated via the antisymmetric with respect to Matsubara frequencies vector component of the normal Green's function  $\bm P_N^M = i[\bm g_N^{y}(\omega)+\tilde{\bm g}_N^{y}(\omega)] = i[\bm g_N^{y}(\omega)-{\bm g}_N^{y}(-\omega)]$. It automatically provides the odd-frequency character of on-site triplets. Turning to the real energies we obtain $\bm P_N^s = i[(\bm g_N^{y,R}(\varepsilon)+ \tilde {\bm g}_N^{y,R}(\varepsilon)] = i[(\bm g_N^{y,R}(\varepsilon)+  {\bm g}_N^{y,R}(-\varepsilon)]$. It is seen from Eq.~(\ref{eq:normal_GFy}) and (\ref{eq:dos}) that this expression determines the symmetric with respect to $\varepsilon$ part of the N\'eel-type spin polarization of the DOS $\bm P_{\bm N}^s \bm N = [P_{\bm N}(\varepsilon)+P_{\bm N}(-\varepsilon)]/2N_F $. Therefore, the N\'eel triplets are generated by the staggered polarization of the normal metal. An analogous term is absent in S/F heterostructures. In the case $h \ll \varepsilon_F$ the spin polarization of normal state is negligible and is considered to be zero in the framework of the quasiclassical approximation. This limit is always relevant for S/F and S/AF heterostructures because $h \lesssim \Delta$ in order to avoid the complete depairing of superconductivity.  

Consequently, the physical mechanisms of triplet generation are different in S/F and S/AF structures. The effect of SOC on the triplets can also be different. It can be seen directly from the solutions of Eqs.~(\ref{eq:eilenberger_triplet}) and (\ref{eq:eilenberger_triplet_F}). Up to the leading order in $h/|i\omega+\mu|$ and $h_R/|i\omega+\mu|$ the on-site N\'eel-type triplet correlations take the form:
\begin{eqnarray}
\bm f^y = {\rm sgn}\omega \left[\frac{i\Delta \bm h}{\mu^2+\omega^2} + \frac{i\Delta [\bm h_R \times (\bm h \times \bm h_R)](3\mu^2-\omega^2)}{(\mu^2+\omega^2)^3}\right]~
\label{eq:triplet_sol_AF}
\end{eqnarray}
It can be compared to the solution for on-site triplet correlation in the S/F thin-film bilayer:
\begin{eqnarray}
\bm f_F^x = -{\rm sgn}\omega \left[\frac{\Delta \bm h}{\omega^2} - \frac{\Delta [\bm h_R \times (\bm h \times \bm h_R)]}{\omega^4}\right].
\label{eq:triplet_sol_F}
\end{eqnarray}
Recall that $\bm f_F^x$ anomalous Green's function accounts for the conventional, not N\'eel, on-site triplet correlations $\bm f_F^x = \bm f_F^{AA} = \bm f_F^{BB} $. 
By comparing Eqs.~(\ref{eq:triplet_sol_AF}) and (\ref{eq:triplet_sol_F}) we see that in S/F heterostructures the SOC weakens triplets, the same is valid for S/AF structures with $\mu \ll \pi T_{c0} $ when we can disregard $\mu$ with respect to $\omega$ in Eq.~(\ref{eq:triplet_sol_AF}). At the same time in S/AF with $\pi T_{c0} \lesssim \mu $ its influence is just opposite - it enhances triplets. In both cases the influence of SOC on triplets is anisotropic. The anisotropy is determined by the vector structure $\bm h_R \times (\bm h \times \bm h_R)$. The influence of SOC on triplets, that is suppression for S/F and enhancement for S/AF bilayers, is maximal when $\bm h_R \perp \bm h$ for all electron trajectories $\bm n_F$, what is realized  for OOP orientation of the magnetization. It is monotonically declines with decreasing $\theta$.

The physical mechanism for the suppression of the triplet correlations with $\bm f \propto \bm h$ by SOC in S/F structures is well-known. The triplets $\bm f = -{\rm sgn}\omega\Delta \bm h/\omega^2$ are obtained from singlets by the process of the singlet-triplet conversion\cite{Buzdin2005}. Then the SOC partially converts these original triplets into the odd in momentum component $\propto \bm h_R \times \bm h$. As a result, the amplitude of initial triplets along $\bm h$ decreases. For dirty systems this process of triplet depairing by SOC is analogous to anisotropic Dyakonov-Perel spin relaxation \cite{Dyakonov1971,Dyakonov1971_2}  because the odd in momentum component $\propto \bm h_R \times \bm h$ is zero after impurity averaging over trajectories and only the reduced triplets along $\bm h$ survive \cite{Bergeret2013,Bergeret2014}. 

On the contrary, the enhancement of triplets by SOC in S/AF structures is caused by the analogous enhancement of the normal state polarization. From Eqs.~(\ref{eq:normal_GFy}) and (\ref{eq:dos}) it follows that up to the leading order in $h/|\varepsilon+\mu|$ and $h_R/|\varepsilon+\mu|$ the N\'eel-type polarization of the normal state DOS along $\bm h$ takes the form 
\begin{eqnarray}
P_{\bm h}^A (\varepsilon) = - P_{\bm h}^B (\varepsilon) = 2N_F \left[ \frac{h}{\varepsilon+\mu} + \frac{h h_R^2 \sin^2 \phi}{(\varepsilon+\mu)^3} \right],
\label{eq:dos_polarization_normal}
\end{eqnarray}
where $h = |\bm h|$ and $\phi$ is the angle between $\bm h$ and $\bm h_R$. It is seen that (i) the absolute value of the polarization is always enhanced by the SOC and (ii) the enhancement is anisotropic. It reaches maximal possible value for all the trajectories for OOP orientation of $\bm h$ because $\bm h_R$ is always in-plane.  It is worth noting that Eq.~(\ref{eq:dos_polarization_normal}) is not applicable at $\mu=0$ because in this limit the most important contribution to the superconducting properties is given by the small energies $\varepsilon \lesssim T_{c0}$, where Eq.~(\ref{eq:dos_polarization_normal}) is not valid because the conditions $h \ll |\varepsilon+\mu|$ and $h_R \ll |\varepsilon+\mu|$ are violated. Therefore, this consideration is not applicable for explanation of the results at $\mu \lesssim \pi T_{c0}$ and we comment on this limit later. 

\begin{figure}[tb]
	\begin{center}
		\includegraphics[width=85mm]{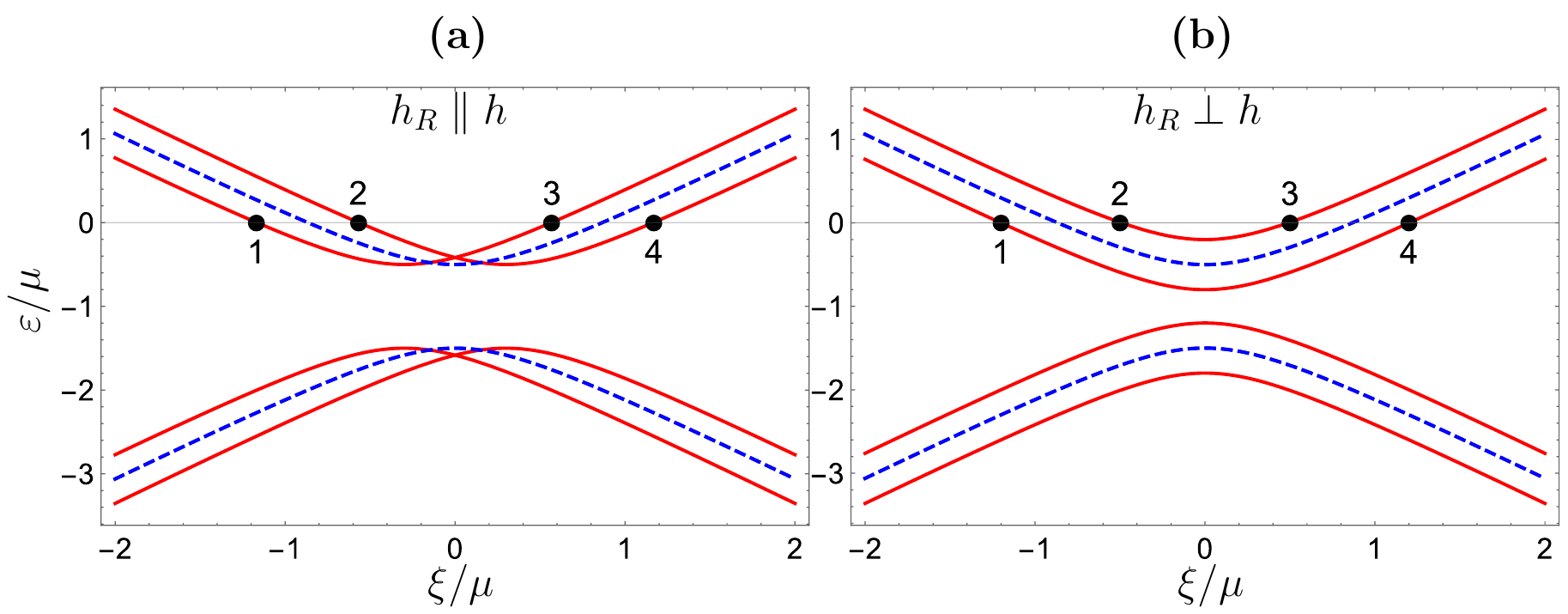}
		\caption{Normal state electron spectra  $\varepsilon(\xi)$ of the S layer in the presence of $\bm h$ and SOC $\bm h_R$, where $\xi = -2t(\cos p_x a + \cos p_y a + \cos p_z a)$. Blue dashed lines correspond to $h_R=0$. Points 1-4 mark the intersection of the spectra with the Fermi surface.}
        \label{fig:spectra}
	\end{center}
\end{figure}

The reason for the described above enhancement is the specific reconstruction of normal state electron spectra under the influence of the SOC. In Fig.~\ref{fig:spectra} we demonstrate the normal state electron spectra of the S layer proximitized by the AF for $\bm h_R \parallel \bm h$ [(a)] and $\bm h_R \perp \bm h$ [(b)]. At $\bm h_R = 0$ the spectra are doubly-degenerated. The eigen states at the Fermi surface for an electron with spin $\sigma = \pm 1$ take the form
\begin{eqnarray}
\left( \begin{array}{c} \hat \psi_{\bm i \sigma}^A \\ \hat \psi_{\bm i \sigma}^B
\end{array}
\right)(\bm p) =  \left( \begin{array}{c} \sqrt{1+\sigma h/\mu} \\ \sqrt{1-\sigma h/\mu}
\end{array}
\right) e^{i \bm p \bm i},
\label{eq:eigen_zero_SOC}
\end{eqnarray}
The distribution of probability density of these states in the lattice strongly oscillates between A and B sites. The total spin polarization at the Fermi level caused by these states is $P_{\bm h}^A = -P_{\bm h}^B = 2 N_F h/\mu$. The nonzero SOC splits the spectra. The splitting is horizontal for $\bm h_R \parallel \bm h$, that is analogous to the conventional action of the Rashba SOC in nonmagnetic metals.  The net spin polarization at the Fermi level is represented by the sum over states $1-4$ in Fig.~\ref{fig:spectra}(a), multiplied by $N_F$. It is again $P_{\bm h}^A = -P_{\bm h}^B = 2 N_F h/\mu$. On the contrary, for $\bm h_R \perp \bm h$ the splitting of the spectra is vertical, which is reminiscent of the Zeeman splitting in conventional metals. Effectively it is described by the opposite shifts of the chemical potentials of the both branches $\mu \to \mu \pm h_R$. The resulting spin polarizations of the states $1-4$ are $P_{1,4}=h/(\mu + h_R)$ and $P_{2,3}=h/(\mu - h_R)$. The total on-site spin polarization is $P_{\bm h}^A = -P_{\bm h}^B = 2N_Fh\mu/(\mu^2-h_R^2)$. Expanding this expression with respect to $h_R/\mu $ we obtain Eq.~(\ref{eq:dos_polarization_normal}) at $\phi=0$ and $\varepsilon=0$. The enhancement of the staggered spin polarization appears as a nonlinear effect of the opposite chemical potential shifts.

If $\mu \lesssim \pi T_c$ the antiferromagnetic gap opens in the superconductor in the vicinity of the normal state Fermi surface. In this case the most important contribution to the pairing correlations is given by the electronic states at the edge of the gap. They correspond to $\xi \approx 0$, what means that the electrons are practically fully localized at one of the sublattices. Consequently, they only feel the magnetization of the corresponding sublattice and behave in the same way as in the ferromagnet. For this reason our results at $\mu \lesssim \pi T_c$ demonstrate the same trends as the corresponding results for S/F structures. 

{\it Summary.}---The effect of Rashba SOC on triplets is studied in S/AF thin film bilayers with fully compensated AFs. A unique effect of anisotropic enhancement of the triplets by SOC is found. It can be experimentally observed via the anisotropy of the superconducting transition. Our analysis highlights the importance of the value of the chemical potential $\mu$ for the physics of S/AF thin-film hybrids. At $\mu \lesssim \pi T_{c0}$ the influence of the SOC on superconducting properties of the S/AF bilayers is the same as for the S/F case - the SOC anisotropically suppresses triplets and enhances $T_c$, and the maximal $T_c$ is reached for OOP orientation of the N\'eel vector. On the contrary, at $\mu > \pi T_{c0}$ the SOC anisotropically enhances triplets and suppresses $T_c$, and $T_c$ is minimal for OOP orientation. The effect can be especially interesting for heterostructures composed of 2D magnets and superconductors because of possibility of external control of the chemical potential. The anisotropy of the superconducting transition opens a perspective for a reciprocal effect, that is the reorientation of the N\'eel vector due to superconductivity. In its turn, this possibility is an important step for further developments in AF-based superconducting spintronics.

\begin{acknowledgments}
The financial support from the Russian
Science Foundation via the RSF project No.22-22-00522 is acknowledged.    
\end{acknowledgments}

\section{Supplemental Material: Two-sublattice formalism of Green's functions}

\label{formalism}

The numerical calculations of the critical temperature are performed in the formalism of the Gor'kov Green's functions in two-sublattice framework\cite{Bobkov2022,Bobkov2023}. The unit cell with two sites in it is chosen as shown in Fig.~1 of the main text. Introducing the two-sublattice Nambu spinor $\check c_{\bm i} = (\hat c_{{\bm i},\uparrow}^A, \hat c_{\bm i,\downarrow}^A, \hat c_{\bm i,\uparrow}^B,\hat c_{\bm i,\downarrow}^B, \hat c_{\bm i,\uparrow}^{A\dagger}, \hat c_{\bm i,\downarrow}^{A\dagger}, \hat c_{\bm i,\uparrow}^{B\dagger}, \hat c_{\bm i,\downarrow}^{B\dagger})^T$ we define the Green's function as $ \check G_{\bm i \bm j}(\tau_1, \tau_2) = -\langle T_\tau \check c_{\bm i}(\tau_1) \check c_{\bm j}^\dagger(\tau_2) \rangle$, 
where $\langle T_\tau ... \rangle$ means  imaginary time-ordered thermal averaging and $\bm i$ is now the sublattice index. The Green's function is a $8 \times 8$ matrix in the direct product of spin, particle-hole and sublattice spaces. Therefore, we  define the Pauli matrices $\bm \sigma = (\sigma_x, \sigma_y, \sigma_z)^T$ in spin space, $\bm \tau = (\tau_x, \tau_y, \tau_z)^T$ in particle-hole space and $\rho = (\rho_x, \rho_y, \rho_z)^T$ in sublattices space. Further we assume the system to be homogeneous along the interface and consider the Fourier-transformed Green's function:
\begin{eqnarray}
\check G(\bm p) = \int d^3 r e^{-i \bm p(\bm i - \bm j)}\check G_{\bm i \bm j},
\label{mixed}
\end{eqnarray}
where the integration is over $\bm i - \bm j$. Then to make the resulting Gor'kov equations simpler it is convenient to define the following  transformed Green's function:
\begin{eqnarray}
\check {\tilde G}(\bm p) = 
\left(
\begin{array}{cc}
1 & 0 \\
0 & -i\sigma_y
\end{array}
\right)_\tau \rho_x e^{\frac{-\displaystyle ip_z a_z \rho_z}{\displaystyle 2}} \times \nonumber \\
\check G(\bm p) 
e^{ \frac{\displaystyle ip_z a_z \rho_z}{\displaystyle 2}}
\left(
\begin{array}{cc}
1 & 0 \\
0 & -i\sigma_y
\end{array}
\right)_\tau ,
\label{unitary}
\end{eqnarray}
where subscript $\tau$ means that the explicit matrix structure corresponds to the particle-hole space. The Gor'kov equation for $\check {\tilde G}(\bm R, \bm p)$ was originally derived in Ref.~\onlinecite{Bobkov2022} and here is generalized to take into account the Rashba SOC. The resulting Gor'kov equation takes the form:
\begin{align}
    [\check H \rho_x  -\xi(\bm p) - \bm h_R(\bm p) \bm \sigma ]\check {\tilde G}=1,
    \label{eq:gorkov_hom} 
\end{align}
\begin{align*}
    \check H  = i \omega_m \tau_z + \mu + \tau_z \check \Delta - \bm h \bm \sigma  \tau_z \rho_z ,
\end{align*}
where $\xi(\bm p) = -2t(\cos p_x a_x + \cos p_y a_y + \cos p_z a_z)$, $\omega_m = \pi T(2m+1)$ is the Matsubara frequency, $\bm h_R(\bm p) = (V_R/2ta) (\bm e_y \times \bm v(\bm p))$ is the effective Rashba pseudomagnetic field seen by an electron moving along the trajectory determined by the velocity $\bm v (\bm p) = d\xi/d \bm p = 2t(\bm a_x \sin[p_x a_x] + \bm a_y \sin[p_y a_y]+ \bm a_z \sin [p_z a_z])$, $\check \Delta = \Delta(\bm R)\tau_+ + \Delta^*(\bm R)\tau_-$ with $\tau_\pm = (\tau_x \pm i \tau_y)/2$. For simplicity we assume $a_x=a_y=a_z=a$. 
The superconducting order parameter $\Delta$ in S is calculated self-consistently:

\begin{align}
    \Delta=-T\sum\limits_{\omega_m} g\int \frac{\mbox{Tr}(\check {\tilde G}(\bm p) \tau_+\sigma_0\rho_x)}{8} \frac{d^3p}{(2\pi)^3},
    \label{eq:self-consistency}
\end{align}
where $g$ is the pairing constant. The critical temperature is calculated from the linearized with respect to $\Delta$ version of Eqs.~(\ref{eq:gorkov_hom}) and (\ref{eq:self-consistency}).

Two-sublattice quasiclassical equations, which were derived in \cite{Bobkov2022}, can also be generalized to take into the SOC. We introduce quasiclassical $\xi$-integrated Green's function:
\begin{eqnarray}
\check g(\bm R, \bm p_F) = -\frac{1}{i \pi} \int \check {\tilde G}(\bm R, \bm p)d\xi,
\label{quasi_green} 
\end{eqnarray}
where $\xi(\bm p) = -2t (\cos p_x a_x+\cos p_y a_y+\cos p_z a_z)$ is the normal state electron dispersion counted from the Fermi energy. Here we also allow for a possible spatial inhomogeneity in the plane of the S/AF interface, where $\bm R$ is the in-plane radius-vector. Performing the standard derivation of the quasiclassical equation \cite{Bobkov2022} one obtains the following Eilenberger equation for the quasiclassical Green's function:
\begin{align}
\left[ \left(i \omega_m \tau_z + \mu + \tau_z \check \Delta(\bm R) - \bm h (\bm R) \bm \sigma \tau_z \rho_z \right)\rho_x-\right. \nonumber \\
\left.- \bm h_R(\bm p_F) \bm \sigma \rho_0, \check g(\bm R,\bm p_F) \right] + i \bm v_F \bm \nabla \check g(\bm R,\bm p_F) = 0 .
\label{eilenberger_ballistic} 
\end{align}

\bibliography{anisotropy}

\end{document}